\documentclass[10pt]{article}
\usepackage{latexsym}
\usepackage{multicol}
\usepackage{amssymb}

\begin{document}
\title{\Large \bf Power laws of the in-degree and out-degree distributions \\
 of complex networks}
\author{{\normalsize \bf Shinji Tanimoto}\\
{\normalsize Department of Mathematics, Kochi Joshi University, Kochi 780-8515, Japan.} \\
{\normalsize (tanimoto@cc.kochi-wu.ac.jp)} 
}
\date{}
\maketitle
\begin{abstract}
A model for directed networks is proposed and power laws for their in-degree and/or out-degree
distributions are derived from the model. 
It is based on the Barab\'asi-Albert model and contains two parameters. The
parameters serve as regulatory factors for the contributions of new nodes through in- and out-degrees.
The model allows a newly added link to connect directly two nodes already present in the network. 
Such a link is called an inner link, while a link attached to a new node is called an outer link. 
Using relationships between
inner and outer links, we investigate power laws for in- and out-degree distributions of directed networks.
This model enables us to predict some interesting features of directed networks; 
in particular, the World Wide Web and the networks of citation and phone-call. 
\end{abstract}
\vspace{0.3cm}
\begin{multicols}{2}
\begin{center}
{\bf\large 1. Introduction}  \\
\end{center}
\indent
\indent
Recent studies of many real-world complex networks
in technological and social systems have revealed that they obey the power law degree 
distribution
\[
p(k) \propto k^{- \gamma}.
\]
Here $p(k)$ is the probability that 
a randomly chosen node (or vertex) has $k$ links (or edges), and the exponent
$\gamma$ is a constant, typically $2 \le \gamma \le 3$. \\
\indent
In order to explain the empirical fact of real networks, Barab\'asi and Albert
proposed a model (BA model) in [3]. It could deduce the degree distribution;
$p(k) \propto k^{-3}$. Their model is thought of as a model for undirected networks. 
Now we are able to find many other network models
from which power law degree distributions with a variety of exponents 
can be deduced (see [6, 8, 11] and references therein).  \\
\indent
The objective of this paper is to propose a model for directed networks whose in-degree and/or
out-degree distributions follow power laws. 
For that purpose, we first review the model of undirected networks introduced in [14]. 
This model modifies the BA model by introducing one parameter
and generates a power law degree distribution.
The parameter regulates the growth rate of a network and its value falls 
between one and two, which reflects the obvious fact that one link has two nodes.  
This simple fact leads us to $2 < \gamma \le 3$. \\
\indent
Unlike the BA model, the proposed model permits old nodes, {\it i.e.}, 
nodes already present in the network,
to connect directly by some of newly added links. We call such links {\it inner links}, 
while links attached to a new node are called {\it outer links}. In BA model 
new links are limited to outer links.  \\
\indent
The Internet and the networks of movie actors or phone-call
can be modeled by our model, since they will grow by acquiring inner links
as well as new nodes and outer links. 
On the other hand, citation networks cannot contain inner links, 
because new links are always attached to new nodes ({\it i.e.}, new papers).  \\
\indent
The model presented for directed networks involves two parameters and
it also allows both inner links and outer links. 
The numbers of those links are determined by two parameters.
The mechanism of connections is based on so-called
`preferential attachment'. It means that the probability that any inner or outer directed link attaches to
an old node is proportional to its in-degree or out-degree.
Considering both incoming and outgoing directions,
two power laws of degree distributions are derived from the model and 
several relationships between both exponents of power laws are discussed. \\
\indent
We will apply these results to the well-known directed networks; 
the World Wide Web, citation networks and phone-call networks. Considering the quantity of
inner links or outer ones, we attempt to predict several rough but 
interesting features of these networks,
in particular, their in- and/or out-degree distributions. \\ 
\indent
A comparison of the ratios of inner links yields a conclusion that 
the WWW can be thought of as an intermediate of 
two extremes; citation networks and phone-call networks. \\
\indent
Our approach to the study of power laws appearing in complex networks
is simple and seems different from others. \\
\begin{center}
{\bf\large 2. Basic model for undirected networks} \\
\end{center}
\indent
\indent
We review the basic model for undirected networks that was introduced in [14] 
so as to facilitate the modeling of directed networks. 
This simple model provides an explanation that degree distributions 
of real-world networks tend to obey power laws with exponents between 2 and 3. 
And it has a desirable feature that old nodes in a network are able to connect directly
through some of newly added links. It is a natural phenomenon in real-world networks, while
the original BA model ([3, 4]) did not take it into account. \\
\indent
In order to explain our model for undirected networks, we divide it into the following five parts (i)--(v). 
\begin{itemize}
\item[{(i)}]  We start with a network having a small number of nodes, say $m_0$ nodes, and some links. 
\item[{(ii)}] New nodes and new links constantly enter the network. 
Whenever a new node enters the network, we advance 
the time step by one and at the same time a constant quantity of links, say $m$ links,
are added to the network. New links are either inner or outer ones.
\end{itemize}
\indent
\indent
Thus new links do not necessarily emanate from the new node. 
The number, $m$, of new links at each time step may be any positive number, 
since we adopt the continuum argument as in [4]. 
The numbers of inner links and outer ones are determined by a proportionality constant 
that is related to preferential attachment. 
The value $m_0$ and the number of the original links are not important, 
since we eventually take the limit with respect to time. \\
\indent
We denote by $v_i$ the $i$th node that enters the network at 
time step $t_i$ and by $k_i$ the degree of the node. 
After $t$ time steps, we obtain a network with $t +m_0$ nodes and $mt$ links plus the original
links. Also notice that the resulting network becomes sparse, since the ratio of links present in the network
at time step $t$ satisfies  
\begin{eqnarray*}
         2mt/[(m_0 + t)(m_0 + t -1)] \rightarrow 0~(t \rightarrow \infty).
\end{eqnarray*}
\indent
Preferential attachment means that the probability, ${\mathcal P}$,
that any node connects to a node $v$ is proportional to the connectivity $k$ of the node $v$,
so that
\begin{eqnarray} \label{1}
{\mathcal P}(k) = \frac {k}{\sum_j {k_j}};  
\end{eqnarray} 
the 'rich-get-richer' phenomenon. \\
\indent
Thus the rate at which a node $v_i$ with $k_i$ links acquires more links is given by
\begin{eqnarray} \label{2}
\frac{dk_i}{dt}= g{\mathcal P}(k_i),    
\end{eqnarray} 
where $g$ is a proportionality constant, which is to be determined later. 
The denominator in (\ref{1}), denoted by 
\begin{eqnarray*}
{\mathcal L}(t)=\sum_j {k_j},
\end{eqnarray*}
is the total of all degrees in the network at time step $t$. In view of (1), 
taking the sums over $i$ in both sides of (2), we have 
\begin{eqnarray} 
\frac{d{\mathcal L}(t)}{dt}= g.    
\end{eqnarray} 
\indent
Since $m$ links are added to the network in each time step and each link is
counted twice in counting degrees, the total degree ${\mathcal L}(t)$ at time step $t$ satisfies
\begin{eqnarray}  
{\mathcal L}(t) \approx 2mt.   
\end{eqnarray}
This implies that the growth rate $g$ of the total degree is not equal to 
$m$ as supposed in [3, 4], but approximately equal to $2m$. The growth rate will be equal to $m$, 
when one completely neglects the degrees of new nodes. On the other hand,
it is equal to $2m$, when one takes a full account of them. \\
\indent
In order to circumvent this discrepancy, we separate the growth of a network into two sides. 
One is the side of old nodes, {\it i.e.}, those already present in the network, and it is governed by 
differential equations (2) and (3). In other words, it is this side that
the 'preferential attachment' rule is applied to.
The other is the side of new nodes and it is prescribed by giving the initial conditions. 
These considerations lead us to the latter parts of the model. 
\begin{itemize}
\item[{(iii)}]
We introduce a parameter $\beta$ into the BA model by $g = \beta m$. 
The parameter serves as a regulatory factor for the growing network.
\item[{(iv)}]
We must set the initial condition for a new node from (3) rather than simply put
$k_i(t_i)=m$. The initial condition depends on $g$ and gives the number of outer links.
\item[{(v)}] The occurrence of inner links are allowed at each time step, 
among $m$ new links, as the rest of outer links.
\end{itemize}
\indent
\indent
It is reasonable to assume that $\beta$ satisfies $1 \le \beta \le 2$ from the argument, and one can
choose any value of $\beta$ in the interval. However,
the condition for $\beta$ will be reconsidered later and it will be replaced by $1 < \beta \le 2$. \\
\indent
Thus Eq. (2), together with (4) and $g =\beta m$, can be written as
\begin{eqnarray} 
\frac{dk_i}{dt} = \beta m \frac{k_i}{2mt} = \beta \frac{k_i}{2t}.   
\end{eqnarray}
As stated in (iv), for a chosen value of $\beta$, we need the corresponding initial condition imposed upon 
a new node $v_i$, which enters the network at time step $t_i$.
Relation (3) implies that $g=\beta m$ is the quantity of degree's increase 
for the side of old nodes within one time step. On the other hand, 
$2m$ is the total quantity of degree's increase. From this we see that the initial condition is given by
\begin{eqnarray}
k_i(t_i) = (2 - \beta)m,    
\end{eqnarray}
for each new node. It gives the number of outer links. \\
\indent
Using (6) and the Dirac delta function $\delta$ ([13]), it would be better to write the rate equation as
\[
       \frac{dk_i}{dt} = \beta \frac{k_i}{2t} + (2-\beta)m\delta (t -t_i) ~~~(t \ge t_i),
\]
instead of (5).
Then it is easy to see that relation (4) naturally follows by its integration and to realize that
the first term on the right hand side is really the growth rate for $v_i$ as an old node.\\
\indent 
Next let us examine the condition for $\beta$ from (6).
It necessarily implies $\beta < 2$ and $\beta \ge 1$. If $\beta \ge 2$, then any proper
networks cannot be generated. When $\beta < 1$, there are outer links more than $m$.
Therefore, the appropriate condition for the parameter $\beta$ is
\begin{eqnarray*}
1 \le \beta < 2.
\end{eqnarray*}
\indent
Moreover, in the case of $1 < \beta < 2$, condition (6)
permits the presence of inner links, which is the content of (v). Namely, 
old nodes themselves present in the network are 
able to connect directly through some of new links. Among $m$ newly added ones, 
there are such inner links as many as
\begin{eqnarray*}
m - (2 - \beta)m = (\beta - 1)m.
\end{eqnarray*}
When $\beta = 1$ as in [3, 4], this model coincides with the original BA model
and all of $m$ new links are connections between a new node and old nodes. \\
\indent
In [14], under the condition $1 \le \beta <2$, it was proved that
this modified BA model ({\it i.e.}, (5) and (6)) generates a network whose
degree distribution follows a power law with exponent 
\begin{eqnarray*}
\gamma = 1 + \frac{2}{\beta}.
\end{eqnarray*}
So we obtain $2 < \gamma \le 3$ from our model.  \\
\indent
The higher the ratio of inner links ({\it i.e.}, $\beta - 1 > 2 -\beta$ and the difference of both sides
 becomes larger), the more $\gamma$ approaches 2.0 from 3.0. 
Table 1 due to [8, p.80] exhibits undirected networks with lower $\gamma$ (or larger $\beta$).
Our intuition tells us that those networks grow by acquiring inner links as well as 
outer ones and new nodes.
\begin{center}
{\footnotesize\sc Table 1}\\
\vspace{1mm}
\begin{tabular}{|c|c|} 
\hline
Network & $\gamma$ ($\beta$) \\
\hline
\hline
movie actors & 2.3 (1.5)\\
\hline
Internet (domain) & 2.2 (1.7)\\
\hline
Internet (router) & 2.3-2.5 (1.3-1.5) \\
\hline
\end{tabular}
\end{center}
\vspace{3mm}
\begin{center}
{\large \bf 3. Directed networks}  \\
\end{center}
\indent
\indent
The framework of the modeling for directed networks with power law degree distributions
is similar to that of undirected networks.
Based on the five procedures (i)--(v) of Section 2 
we deduce two degree distributions for a directed network, considering
both incoming and outgoing directions. \\
\indent
As before,
at each time step, one new node and $m$ new directed links are added to the network. This model
also allows presence of inner links connecting directly old nodes that already exist in the network.
The mechanism of preferential attachment is used when
inner or outer directed links attach to old nodes 
is proportional to their in-degrees or out-degrees. \\
\indent
Let us denote by $k_i$ the in-degree of $i$th node $v_i$.
First we consider the in-degree distribution and assume that preferential attachment can be applied to 
in-degrees. 
If we denote by ${\mathcal L}(t)$ the total in-degree of 
the network at time step $t$, then ${\mathcal L}(t) \approx mt$ instead of (4).   
Hence, using (iii) and (iv) of Section 2, the rate equation (5) and the initial condition (6) should be 
replaced by
\begin{eqnarray}
\frac{dk_i}{dt} = \alpha m \frac{k_i}{mt} = \alpha \frac{k_i}{t}, ~ ~ ~k_i(t_i) = (1 - \alpha)m, 
\end{eqnarray}
respectively. 
Here the parameter $\alpha$ regulates the growth rate of network's in-degrees and $0 < \alpha < 1$. 
Remark that $\alpha m$ is a proportionality constant 
that is related to preferential attachment. It is the quantity of in-degree's increase 
within one time step for the side of old nodes.
Alternatively, using the delta function, one can write them in a single equation as
\[
       \frac{dk_i}{dt} = \alpha \frac{k_i}{t} + (1-\alpha)m\delta (t -t_i) ~~~(t \ge t_i).
\]
\indent
Employing a similar argument to the one in [14], we get from (7)
\begin{eqnarray*}
p(k) \propto k^{-(1+1/\alpha)}
\end{eqnarray*}
for the in-degree distribution. 
In what follows the derivation of this is given for the sake of completeness. 
For another proof using the master equations see [8, 14]. \\
\indent
The solution of Eq. (7) is given by  
\begin{eqnarray*}
k_i(t) = (1 - \alpha)m\left( \frac{t}{t_i}\right)^{\alpha}~~~(t \ge t_i),
\end{eqnarray*}
where $t_i$ is the time when a node $v_i$ enters the network. 
It is assumed that the time interval $s=t_{i+1}-t_i$ (for each $i$) is a constant, usually $s=1$.\\
\indent
In order to obtain the in-degree distribution for our model, we could adopt the method employed in [3, 4].
However, we will use the following alternative approach, because it is easier to follow and, moreover,
it gives the same asymptotic behavior for the in-degree distribution as 
the method in [3] or [4]. \\
\indent
To deduce the probability $p(k)$ that a randomly chosen node $v$ has an in-degree $k$, 
it suffices to estimate the number of time steps $t_i$'s for nodes $v_i$'s, 
which satisfy the inequalities
\begin{eqnarray} 
         k-1 < k_i(t) = (1 - \alpha)m\left(\frac{t}{t_i}\right)^{\alpha} \le k.    
\end{eqnarray}
This can be written as
\begin{eqnarray*}
         t\left[\frac{k}{(1 - \alpha)m}\right]^{-1/\alpha} \le t_i < 
	 t \left[\frac{k-1}{(1 - \alpha)m}\right]^{-1/\alpha}.
\end{eqnarray*}
In the case of $s = 1$, the number of nodes satisfying (8) is
equal to the difference of both sides of the above inequalities:
\begin{eqnarray*}
t\{(1-\alpha)m\}^{1/ \alpha}\{(1 - 1/k)^{-1/\alpha}-1\}k^{-1/\alpha}.
\end{eqnarray*} 
Making use of $(1-1/k)^{-1/\alpha} \approx 1+{\alpha}^{-1}/k$ for large $k$,
we see that the number of nodes satisfying (8) is approximated by 
\begin{eqnarray*}
\alpha^{-1}t\{(1-\alpha)m\}^{1/ \alpha}k^{-1-1/\alpha}.
\end{eqnarray*}
Since at time $t$ there exist $t + m_0$ nodes, the probability $p(k)$ is thus given by
\begin{eqnarray*}
\frac{\alpha^{-1}t\{(1-\alpha)m \}^{1/ \alpha}}{t + m_0}k^{-1-1/\alpha}.
\end{eqnarray*}
In the limit $t \to \infty$ we conclude that
\begin{eqnarray*}
p(k) \propto k^{-(1+1/\alpha)},
\end{eqnarray*}
for all large $k$, {\it i.e.}, $\gamma_{\rm in} = 1 + 1/\alpha$.\\
\indent
Next we assume that preferential attachment can be applied to out-degrees and denote by $\ell_i$ 
the out-degree of the $i$th node. 
Solving a similar equation to (7) for $\ell_i$ 
\begin{eqnarray}
\frac{d{\ell}_i}{dt} = \beta m \frac{{\ell}_i}{mt} = \beta \frac{{\ell}_i}{t}, ~ ~ ~{\ell}_i(t_i) = (1 - \beta)m, 
\end{eqnarray}
leads to another power law for the out-degree distribution
\begin{eqnarray*}
p(\ell) \propto {\ell}^{-(1+1/\beta)}.
\end{eqnarray*}
Here the parameter $\beta$ regulates the growth rate of network's out-degrees 
and $0 < \beta < 1$. \\
\indent
In summary the in- and out-degree distributions of directed networks follow the power laws with
exponents
\begin{eqnarray}
\gamma_{\rm in} = 1 + \frac{1}{\alpha}, ~~~ \gamma_{\rm out} = 1 + \frac{1}{\beta}.
\end{eqnarray}
\indent
Using the initial conditions of (7) and (9), among $m$ newly added links, there are outer links 
as many as 
\[
(2 - \alpha - \beta)m,
\]
while there are
\[ (\alpha + \beta - 1)m
\]
inner links, {\it i.e.}, not attached to a new node, as stated in (v) of Section 2. 
When $\alpha + \beta = 2$, note that any proper networks are not generated.
Both numbers of links entail the inequalities
\begin{eqnarray}
1 \le \alpha + \beta < 2.
\end{eqnarray}
\indent
When $\alpha \lessapprox 1$ ({\it i.e.}, $\alpha$ is close to and less than 1), (10) implies that
the higher the ratio of inner links (namely, $\alpha + \beta - 1 > 2 -\alpha -\beta$ and the difference
between both sides becomes larger), 
the more the exponent $\gamma_{\rm out}$ approaches 2.0 from 3.0. 
This remark will be used in the next section. 
\vspace{2mm}
\begin{center}
{\large \bf 4. The WWW and the networks of citation and phone-call}  \\
\end{center}
\indent
\indent
In this section we deal with three kinds of directed networks with power law degree distributions;
the World Wide Web, citation networks and phone-call networks.  
Table 2 due to [5, p.72] or [8, p.80] exhibits exponents of their
power law in- and/or out-degree distributions. \\
\begin{center}
{\footnotesize\sc Table 2}\\
\vspace{1mm}
\begin{tabular}{|c|c|c|} 
\hline
Network (reference) & $\gamma_{\rm in}$ ($\alpha$) & $\gamma_{\rm out}$ ($\beta$) \\
\hline
\hline
WWW ([9]) & 2.1 (0.9) & 2.38 (0.7) \\
\hline
WWW ([2]) & 2.1 (0.9) & 2.45 (0.7) \\
\hline
WWW ([7]) & 2.1 (0.9) & 2.72 (0.6) \\
\hline
citation ([12]) & 3.0 (0.5) & --- (0) \\  
\hline
citation ([15]) & 2.0 (1.0) & --- (0) \\
\hline
phone-call ([1]) & 2.1 (0.9) &  2.1 (0.9) \\
\hline
\end{tabular}
\end{center}
\vspace{2mm}
\indent
\indent
We apply the results of the previous section separately 
to each of those networks. Using Table 2, we compare 
the power law exponents of their in- and/or out-degree distributions
with those that are deduced from the model. 
For that purpose we examine the quantity or ratio of inner or outer links in new directed links
and the initial conditions for incoming and/or outgoing links. \\
\indent
It will be shown that the ratios of inner links of the WWW, a citation network and a phone-call
network are approximately equal to 0.5, 0 and 1.0, respectively. 
A conclusion drawn from the observations is that 
the WWW can be thought of as an intermediate of 
two extremes; citation networks and phone-call networks. \\ 
\begin{center}
A. The WWW \\
\end{center}
\indent
\indent
The World Wide Web (WWW) is a directed network with nodes (Web pages) and hyperlinks. 
Since its first appearance in the early 1990s, the WWW has been extensively studied, 
beginning with seminal works [2, 3, 7] among others.
It is assumed that
preferential attachment can be applied to both in- and out-degrees. So,
the WWW has power law in- and out-degree distributions and other interesting properties. \\
\indent
Due to [8] or [10], in May 1999 the WWW contained $203 \times 10^6$ nodes 
and $1466 \times 10^6$ hyperlinks,
and in October 1999 there were $271 \times 10^6$ nodes and $2130 \times 10^6$ hyperlinks. 
This data suggests a large amount of inner links. \\
\indent
In the WWW the initial incoming links of most nodes seem relatively rare.
So, setting $\alpha \lessapprox 1$ in (7), we obtain the inequality
\[
\gamma_{\rm in} \gtrapprox 2
\] 
from (10). 
The model explains the reason why the exponents $\gamma_{\rm in}$ 
of the in-degree distributions for the WWW are close to 2.0 as Table 2 shows. \\
\indent
On the other hand, most new nodes of the WWW have more outgoing links
than incoming ones. So, using the initial conditions in (7) and (9), we have 
$1 - \alpha < 1 - \beta$, and hence generally the inequality
\[
\gamma_{\rm in} < \gamma_{\rm out}
\]
also holds from (10). Table 2 tells us that this inequality together with (11) is satisfied 
in the WWW. \\
\indent
The assumption that the number of inner links is larger than that of outer ones is equivalent to the
inequality
\[
\gamma_{\rm out} < 3.
\] 
This immediately follows from the fact that two relations
\[
\alpha + \beta - 1 \ge 2 -\alpha -\beta,~~~ \alpha \lessapprox 1,
\]
yield $\beta > 0.5$ and $\gamma_{\rm out} < 3$, and vice versa. 
Table 2 suggests that such an assumption is effective and
that $\alpha + \beta \gtrapprox 1.5$ holds in the WWW, which implies
$(\alpha + \beta -1)m \ge (2 - \alpha - \beta)m$. Therefore, 
the number of inner links, among $m$ new directed links, tends to be larger than that of outer links,
but the difference does not become so large. \\
\begin{center}
B. Citation network \\
\end{center}
\indent
\indent
A similar argument may be employed in a citation network of academic papers.
The studies of citation networks have a long history than the WWW (see [6, 11]). 
Suppose 
preferential attachment can be applied to cited papers; another 'rich-get-richer' phenomenon. 
For the same reason as above, we can set $\alpha \lessapprox 1$ in (7) for the initial condition of new papers.
Thus the in-degree distribution for a citation network turns out to obey a power law with
exponent $\gamma_{\rm in} \gtrapprox 2$. This result is in favor of [15] rather than 
[12]; $\gamma_{\rm in} \approx 3$. \\
\indent
On the other hand, we must set $\beta \gtrapprox 0$, 
since there are no inner links in a citation network. However,
we cannot apply (10) for $\gamma_{\rm out}$, because we do not know whether
preferential attachment is valid for out-degrees in a citation network. \\ 
\begin{center}
C. Phone-call network \\
\end{center}
\indent
\indent
As for the in-degree distribution of a phone-call network, Table 2 also exhibits the exponent 
with lower $\gamma_{\rm in}$ and hence $\alpha \lessapprox 1$. 
The reason is the same as a citation network.  \\
\indent
In contrast to a citation network or the WWW, however, we guess that 
there are much more inner links
than outer ones in a phone-call network. It implies
$\beta \lessapprox 1$ from the remark at the end of Section 3 and this can be verified by Table 2, too.
If preferential attachment is also valid for out-degrees in a phone-call network, then
the out-degree distribution obeys a power law and its exponent
$\gamma_{\rm out}$ tends to be close to 2.0, as reported in an early paper [1]. \\
\begin{center}
{\bf\large References}
\end{center}
\begin{itemize}
\item[{[1]}] W. Aiello, F. Chung and L. Lu, Proceedings of the 32nd ACM Symposium on the Theory of
Computing, ACM Press, New York, pp. 271--180, 2000.
\vspace{-3mm}
\item[{[2]}] R. Albert, H. Jeong and A.-L. Barab\'asi, Diameter of the World-Wide Web, {\it Nature} {\bf 400}
pp. 130--131, 1999.
\vspace{-3mm}
\item[{[3]}] A.-L. Barab\'asi and R. Albert, Emergence of scaling in random networks, 
{\it Science} {\bf 286} pp. 509--512, 1999.
\vspace{-3mm}
\item[{[4]}]  A.-L. Barab\'asi, R. Albert and H. Jeong, 
Mean-field theory for scale-free random networks, {\it Physica A} {\bf 272} pp. 173--187, 1999.
\vspace{-3mm}
\item[{[5]}]  A.-L. Barab\'asi, in {\it Handbook of Graphs and Networks}, 
ed.~S. Bornholdt and H. G. Schuster, Wiley-VCH, New York, pp. 69--84, 2003.
\vspace{-3mm}
\item[{[6]}]  S. Boccaletti, V. Latora, Y. Moreno,  M. Chavez and D.-U. Hwang, 
Complex networks: Structure and dynamics,
{\it Physics Reports} {\bf 424} pp. 175--308, 2006.   
\vspace{-3mm}
\item[{[7]}] A. Broder, R. Kumar, F. Maghoul, P. Raghavan, S. Rajalopagan, R. Stata, A. Tomkins,
and J. Wiener, Graph structure in the web, {\it Computer Networks} {\bf 33} pp. 309--320, 2000.
\vspace{-3mm}
\item[{[8]}]  S. N. Dorogovtsev and J. F. F. Mendes,
{\it Evolution of Networks: From Biological Nets to the Internet and WWW}, 
Oxford Univ. Press, Oxford, 2003.
\vspace{-3mm}
\item[{[9]}] R. Kumar, P. Raghavan, S. Rajalopagan and A. Tomkins, 
Proceedings of the 9th ACM Symposium on Principles of Database Systems, 1999.
\vspace{-3mm}
\item[{[10]}] J. Liu, Y. Dang, Z. Wang and T. Zhou, 
Relationship between the in-degree and out-degree of WWW,
{\it Physica} A {\bf 371} pp. 861--869, 2006.
\vspace{-3mm}
\item[{[11]}]  M. E. J. Newman, The structure and function of complex networks,
{\it SIAM Review} {\bf 45} pp. 167--256, 2003.     
\vspace{-3mm}
\item[{[12]}] S. Redner, How popular is your paper? An empirical study of the citation
distribution, {\it Eur. Phys. J.} B {\bf 4} pp. 131--134, 1998.
\vspace{-3mm}
\item[{[13]}]  L. I. Schiff, {\it Quantum Mechanics}, McGraw-Hill, 1968.
\vspace{-3mm}
\item[{[14]}] S. Tanimoto, A remark on the BA model of scale-free networks,
arXiv:physics/09021570v3, 2009.
\vspace{-3mm}
\item[{[15]}] A. V\'azquez, Statistics of citation networks, arXiv:cond-mat/0105031, 2001.
\end{itemize}
\end{multicols}
\end{document}